\documentclass[10pt,conference]{IEEEtran}
\ifCLASSINFOpdf
\else
\fi
\usepackage{amsmath}
\usepackage{amsmath}
\usepackage{amssymb}

\newtheorem{Th}{Theorem}
\newtheorem{Lem}{Lemma}

\begin{document}
%
\title{On the Courtade-Kumar conjecture for certain classes of Boolean functions}
%
%
%

\author{Septimia~Sarbu, septimia.sarbu@gmail.com}
\maketitle

\begin{abstract}
We prove the Courtade-Kumar conjecture, for certain classes of $n$-dimensional Boolean functions, $\forall n\geq 2$ and for all values of the error probability of the binary symmetric channel, $\forall 0 \leq p \leq \frac{1}{2}$. Let $\mathbf{X}=[X_1 \enskip \ldots \enskip X_n]$ be a vector of independent and identically distributed Bernoulli$(\frac{1}{2})$ random variables, which are the input to a memoryless binary symmetric channel, with the error probability in the interval $0 \leq p \leq \frac{1}{2}$, and $\mathbf{Y}=[Y_1 \enskip \ldots \enskip Y_n]$ the corresponding output. Let $f:\{0,1\}^n \rightarrow \{0,1\}$ be an $n$-dimensional Boolean function. Then, the Courtade-Kumar conjecture states that the mutual information $\operatorname{MI}(f(\mathbf{X}),\mathbf{Y}) \leq 1-\operatorname{H}(p)$, where $\operatorname{H}(p)$ is the binary entropy function.
\end{abstract}


%
\IEEEpeerreviewmaketitle

\section{Introduction}
%
%
%
%
A recent information-theoretic conjecture, termed the Courtade-Kumar conjecture, was stated in \cite{Courtade2014} and gives the upper bound on the mutual information between a Boolean function of a random vector of inputs to a memoryless binary symmetric channel and the vector of the outputs. The mutual information is computed between a Boolean function of $n$ independent and identically distributed Bernoulli random variables, with success probability, $q=\frac{1}{2}$, and the output of a memoryless binary symmetric channel, with error probability, $0 \leq p \leq \frac{1}{2}$, when this vector of Bernoulli random variables is passed as its input. The conjecture states that this upper bound is equal to $1-\operatorname{H}(p)$, where $\operatorname{H}(p)$ denotes the binary entropy function. Several proofs have appeared in the literature, for different settings of this conjecture, but the most general case has remained unsolved. We bring further contributions to this effort. Using Karamata's theorem \cite{Karamata1932}, we prove the Courtade-Kumar conjecture \cite{Courtade2014}, for certain classes of Boolean functions, $\forall n\geq 2$ and $\forall 0 \leq p \leq \frac{1}{2}$. These functions represent particular subclasses of lex functions, as introduced by Kumar and Courtade in \cite{Kumar2013}. In the context of this conjecture, Karamata's theorem has been used in an earlier version of the preprint \cite{Kindler2016}, which extends the conjecture to the continous case. The generalization of Karamata's theorem, named Schur convexity, has been employed in \cite{Kumar2013}.

Our paper is structured as follows: we start the introductory section with the prior results obtained so far in the literature, in the effort to solve the Courtade-Kumar conjecture. We end this section with our contributions. The essence of this paper, the proof of the Courtade-Kumar conjecture for particular classes of Boolean functions, for any dimension $n\geq 2$ and any error probability $0 \leq p \leq \frac{1}{2}$, is given in Section \ref{ProofCK}. We present the conclusions of this study in Section \ref{Conc}. 
\subsection{Prior work related to the Courtade-Kumar conjecture}
The proofs that have made the most progress towards solving the Courtade-Kumar conjecture are \cite{Ordentlich2016}, \cite{Alex2016}. The authors of \cite{Ordentlich2016} employ Fourier analysis and the hypercontractivity theorem to prove the bound stated in their Theorem $1$, in the case of balanced Boolean functions and $p$ in the range $\frac{1}{2} \cdot \left( 1-\frac{1}{\sqrt{3}}\right) \leq p \leq \frac{1}{2}$: $\operatorname{MI}(f(\mathbf{X}),\mathbf{Y}) \leq \frac{\log{(e)}}{2} \cdot (1-2\cdot p)^2 + 9 \cdot \left(1-\frac{\log{(e)}}{2}\right) \cdot (1-2\cdot p)^4 $. They show that this new bound performs better than the previously established bound of $(1-2 \cdot p)^2$ of \cite{Erkip1996}, in the case of $\frac{1}{3} \leq p \leq \frac{1}{2}$. In Corollary $1$, they prove that the Courtade-Kumar conjecture holds for the dictatorship function, as a special case of equiprobable Boolean functions, when $p \rightarrow \frac{1}{2}$. This region is termed the noise interval $p \in [\frac{1}{2}-\overline{p_n} \enskip \frac{1}{2}]$, where $\overline{p_n}$ is defined as $\overline{p_n}=\frac{1}{4} \cdot 2^{-n}$. Related to this result, in Theorem $1.15$, the author of \cite{Alex2016} proves that the Courtade-Kumar conjecture holds for high noise, that is $\operatorname{MI}(f(\mathbf{X}),\mathbf{Y}) \leq 1-\operatorname{H}(p)$ holds for any Boolean function and for any noise $\epsilon \geq 0$, such that $(1-2 \cdot \epsilon)^2 \leq \delta$ $\Leftrightarrow  \frac{1}{2}-\frac{\sqrt{\delta}}{2} \leq \epsilon \leq \frac{1}{2}+\frac{\sqrt{\delta}}{2}$, where $\delta > 0$ is a constant of small value. The author of \cite{Alex2016} provides an improvement of Theorem $1$ derived by Wyner and Ziv in \cite{Wyner1973}, known as Mrs. Gerber's Lemma, which was employed in \cite{Erkip1996}, for the proof of Theorem $4$. This strenghtening of Mrs. Gerber's Lemma is employed in the proof of the Courtade-Kumar conjecture for high noise \cite{Alex2016}. 

An extension of the Courtade-Kumar conjecture to two $n-$dimensional Boolean functions, is hypothesized to hold in \cite{Anantharam2013}, termed Conjecture $3$. It states that, for any Boolean functions $f,g:\{0,1\}^n \rightarrow \{0,1\}$, the mutual information $\operatorname{MI}(f(\mathbf{X}),g(\mathbf{Y})) \leq 1-\operatorname{H}(p)$. For several specific cases of the joint probability mass function of the binary random variables $f(\mathbf{X})$ and $g(\mathbf{Y})$, the authors analytically prove another conjecture, termed Conjecture $4$,  which implies Conjecture $3$. A similar form of Conjecture $4$ of \cite{Anantharam2013} is analytically proved in \cite{Calmon2014}, in a more general context than that of the results of \cite{Anantharam2013}. In section V of \cite{Calmon2014}, the authors prove that the mutual information $\operatorname{MI}(B,\hat{B}) \leq 1-\operatorname{H}(p)$, for Boolean functions, $B=f(\mathbf{X})$ and $\hat{B}=g(\mathbf{Y})$, an estimator of $\mathbf{Y}$, with fixed mean $\mathbb{E}(B)=\mathbb{E}(\hat{B})=a$ and $\mathbb{P}(B=\hat{B}=0) \geq a^2$. Conjecture $3$ of \cite{Anantharam2013} is proved to hold in \cite{Pichler2016}. The Courtade-Kumar conjecture is generalized to continuous random variables in the preprint \cite{Kindler2016}. The function $f$ takes as input $n-$dimensional real vectors, when they are correlated Gaussian random vectors and when they are correlated random vectors from the unit sphere. As output, the function produces values from the set $\{0,1\}$. 

\subsection{Our contributions}
\begin{Th}
\label{ThCK}
Let $\mathbf{X}=[X_1 \enskip X_2 \enskip \ldots\enskip X_n]$ be an $n$-dimensional random vector of independent and identically distributed Bernoulli$(\frac{1}{2})$ random variables and $\mathbf{Y}=[Y_1 \enskip Y_2 \enskip \ldots \enskip Y_n]$ the result of sending $\mathbf{X}$ through a discrete memoryless binary symmetric channel, without feedback and with the error probability $0 \leq p \leq \frac{1}{2}$. Let $f:\{0,1\}^n \rightarrow \{0,1\}$ be an $n$-dimensional Boolean function, which has any of the following properties: $(1)$ for any $\mathbf{X}^{(i)} \in \{0,1\}^n$, $f(\mathbf{X^{(i)}})=1$,$f(\mathbf{X})=0, \forall \mathbf{X} \in \{0,1 \}^n, \mathbf{X} \neq \mathbf{X^{(i)}}$; $(2)$ for any $\mathbf{X^{(i)}} \in \{0,1\}^n$, $f(\mathbf{X^{(i)}})=0, f(\mathbf{X})=1, \forall \mathbf{X} \in \{0,1 \}^n, \mathbf{X} \neq \mathbf{X^{(i)}}$; $(3)$ $\mathbf{X^{(i)}}=[\mathbf{X_r} \enskip \mathbf{X_{n-r}^{(i)}}]$, $\forall \mathbf{X_{n-r}^{(i)}}  \in \{0,1\}^{n-r}$, that is $ i \in \{1,2, \ldots, 2^{n-r} \}$, $\forall r \in \{1,2,\ldots,n-1 \}$, $f(\mathbf{X^{(i)}})=1$, $f(\mathbf{X})=0, \forall \mathbf{X} \in \{0,1 \}^n, \mathbf{X} \neq \mathbf{X}^{(i)}$; $(4)$ $\mathbf{X^{(i)}}=[\mathbf{X_r} \enskip \mathbf{X_{n-r}^{(i)}}]$, $\forall \mathbf{X_{n-r}^{(i)}}  \in \{0,1\}^{n-r}$, that is $ i \in \{1,2, \ldots, 2^{n-r} \}$, $\forall r \in \{1,2,\ldots,n-1 \}$ $f(\mathbf{X^{(i)}})=0$, $f(\mathbf{X})=1, \forall \mathbf{X} \in \{0,1 \}^n, \mathbf{X} \neq \mathbf{X}^{(i)}$. Let $\operatorname{H}(p)$ denote the binary entropy function. Then, $\displaystyle{\quad \quad \operatorname{MI}(f(\mathbf{X}),\mathbf{Y}) \leq 1-\operatorname{H}(p), \forall n \geq 2,\forall 0 \leq p \leq \frac{1}{2}}$.
\end{Th}

\section{Proof of the Courtade-Kumar conjecture, for certain classes of $n$-dimensional Boolean functions, $\forall n\geq 2$ and $\forall 0\leq p \leq \frac{1}{2}$}
\label{ProofCK}
\begin{Lem}
\label{Lem0}
For any $ k \in \{1,2, \ldots,n \}$, let $\mathbf{Y}=[y_1 \enskip y_2 \ldots y_k] \in \{0,1\}^k$ be fixed and $\mathbf{X^{(i)}}=[x_1^{(i)} \enskip x_2^{(i)} \ldots x_k^{(i)}] \in \{0,1\}^k$ range over all the $2^k$ possible values. Then, the following identy holds
$\sum_{i=1}^{2^{k}} \operatorname{p}(\mathbf{Y},\mathbf{X^{(i)}})=\frac{1}{2^{k}}$.%
\end{Lem}
\begin{IEEEproof}
$\mathbf{X^{(i)}}$ ranges from $[0 \enskip 0 \ldots 0]$ to $[1 \enskip 1 \ldots 1]$. For any fixed $\mathbf{Y}$, there is one $\mathbf{X^{(i)}}$, such that $\mathbf{X^{(i)}}=\mathbf{Y}$. There are $\displaystyle{\binom{k}{1}}$ number of vectors $\mathbf{X^{(i)}}$ that differ from $\mathbf{Y}$ in one position. There are $\displaystyle{\binom{k}{j}}$ number of vectors $\mathbf{X^{(i)}}$ that differ from $\mathbf{Y}$ in $j$ positions. As a result, the summation of the joint probabilities becomes $\displaystyle{\sum_{i=1}^{2^{k}} \operatorname{p}(\mathbf{Y},\mathbf{X^{(i)}})=\sum_{i=1}^{2^{k}} \prod_{j=1}^{k} \operatorname{p}(y_j,x_j^{(i)})= \sum_{r=0}^{k} \binom{k}{r} \cdot}$ $\displaystyle{\frac{(1-p)^{k-r} \cdot p^r}{2^k}=\frac{1}{2^k}}$.
\end{IEEEproof}

\subsection{Boolean functions from the classes $1$ and $2$ of Theorem \ref{ThCK}} 
In order to apply Karamata's inequality \cite{Karamata1932}, we need to transform the mutual information into an algebraic expresion. To this end, we employ concepts from probability mass functions of transformations of random variables $[$ Ch $5$, section $6$ of \cite{MGB1974} $]$. Let $\mathbf{X},\mathbf{Y}$ be two $n-$dimensional discrete random vectors, with ensembles $\mathcal{E}_\mathbf{X}$, $\mathcal{E}_\mathbf{Y}$, $Z$ a discrete random variable, with ensemble $\mathcal{E}_Z$, and an $n$-dimensional function $f$, such that $Z=f(\mathbf{X})$. Let $\mathbf{T},\mathbf{U}$  be two random vectors and $g$ be a multidimensional function, such that $\mathbf{T}=g_1(\mathbf{X},\mathbf{Y})=\mathbf{Y}$, $\mathbf{U}=g_2(\mathbf{X},\mathbf{Y})=\mathbf{X}$ and $Z=g_3(\mathbf{X},\mathbf{Y})=f(\mathbf{X})$. 
\begin{IEEEeqnarray}{Rl}
&\operatorname{p_{\mathbf{TU}Z}}(\mathbf{t},\mathbf{u},z)=\sum_{\begin{matrix} \scriptstyle{\mathbf{x} \in \mathcal{E}_\mathbf{X}, \mathbf{y} \in \mathcal{E}_\mathbf{Y},} \\ \scriptstyle{g_1(\mathbf{x}, \mathbf{y})=\mathbf{t}}, \end{matrix}}  \sum_{\begin{matrix} \scriptstyle{g_2(\mathbf{x},\mathbf{y})=\mathbf{u}} \\ \scriptstyle{g_3(\mathbf{x},\mathbf{y})=z}  \end{matrix}} \operatorname{p_{\mathbf{XY}}}(\mathbf{x},\mathbf{y}) \IEEEnonumber \\
&\operatorname{p_{\mathbf{Y}Z}}(\mathbf{y},1)=\sum_{\mathbf{u} \in \mathcal{E}_\mathbf{U}} \operatorname{p_{\mathbf{TU}Z}}(\mathbf{t},\mathbf{u},1)=\sum_{\mathbf{u} \in \mathcal{E}_\mathbf{U},1=f(\mathbf{u})} \operatorname{p_{\mathbf{XY}}}(\mathbf{u},\mathbf{t}) = \IEEEnonumber \\
&=\sum_{\mathbf{x} \in \mathcal{E}_\mathbf{X}, 1=f(\mathbf{x})} \operatorname{p_{\mathbf{XY}}}(\mathbf{x},\mathbf{y}); \operatorname{p_{\mathbf{Y}Z}}(\mathbf{y},0)=\operatorname{p_{\mathbf{Y}}}(\mathbf{y})-\operatorname{p_{\mathbf{Y}Z}}(\mathbf{y},1). \IEEEnonumber
\end{IEEEeqnarray}
Let $N_0$, $N_1$, $\{\mathbf{x_{i}^{(0)}} \}$ and $\{\mathbf{x_{k}^{(1)}} \}$, such that $f(\mathbf{x_{i}^{(0)}})=0$ and $f(\mathbf{x_{k}^{(1)}})=1$, $\forall i \in \{1,2, \ldots N_0 \}$, $\forall k \in \{1,2, \ldots N_1 \}$. For the first class of functions, $N_1=1$, $N_0=2^n-1$. Then, $\operatorname{p_{\mathbf{Y}Z}}(\mathbf{y},1)= \operatorname{p_{\mathbf{XY}}}(\mathbf{x_{1}^{(1)}},\mathbf{y})$, $\operatorname{p_{\mathbf{Y}Z}}(\mathbf{y},0)=\frac{1}{2^n}-\operatorname{p_{\mathbf{Y}Z}}(\mathbf{y},1)$, $\forall \mathbf{y} \in \mathcal{E}_\mathbf{Y}=\{0,1\}^n$. For any $\mathbf{x_{1}^{(1)}} \in \{0,1 \}^n$, there exists: one vector, that is $m_0=1$, $\mathbf{y_{i_0}} \in \{0,1 \}^n$, such that $\mathbf{y_{i_0}}=\mathbf{x_{1}^{(1)}}$, a number $m_1=\binom{n}{1}$ of the vectors $(\mathbf{y_{i_1}})$, $\forall i_1 \in \{m_0+1,m_0+2,\ldots,m_0+m_1 \}$, such that $(\mathbf{y_{i_1}})$ differ from $\mathbf{x_{1}^{(1)}}$ in one position and a number $m_k=\binom{n}{k}$ of the vectors $(\mathbf{y_{i_k}})$, $\forall i_k \in \{(m_0+\ldots+m_{k-1})+1,(m_0+\ldots+m_{k-1})+2,\ldots,(m_0+\ldots+m_{k-1})+m_k \}$, such that $(\mathbf{y_{i_k}})$ differ from $\mathbf{x_{1}^{(1)}}$ in $k$ positions, $\forall k \in \{0,1,2, \ldots n\}$.
\begin{IEEEeqnarray}{Rl}
&\operatorname{p_{\mathbf{Y}Z}}(\mathbf{y_{i_k}},1)=\frac{(1-p)^{n-k} \cdot p^k}{2^n},  \operatorname{p_{\mathbf{Y}Z}}(\mathbf{y_{i_k}},0)=\frac{1}{2^n}-\IEEEnonumber \\
&\operatorname{p_{\mathbf{Y}Z}}(\mathbf{y_{i_k}},1) , \forall i_k \in \{(m_0+\ldots+m_{k-1})+1,\ldots, \IEEEnonumber \\
&(m_0+\ldots+m_{k-1})+m_k \}, m_{k}=\binom{n}{k}, \forall k \in \{0,1,\ldots,n \}. \IEEEnonumber \\
&\operatorname{p_Z}(1)=\sum_{i=1}^{2^n} \operatorname{p_{\mathbf{Y}Z}}(\mathbf{y_{i}},1)=\frac{1}{2^n}, \operatorname{p_Z}(0)=1-\operatorname{p_Z}(1)=\frac{2^n-1}{2^n}. \IEEEnonumber%
\end{IEEEeqnarray}
\begin{IEEEeqnarray}{Rl}
\label{MIdef1}
&\operatorname{MI}(\mathbf{Y},Z)=\sum_{\mathbf{y}} \sum_z \operatorname{p_{\mathbf{Y}Z}}(\mathbf{y},z) \cdot \log{\frac{\operatorname{p_{\mathbf{Y}Z}}(\mathbf{y},z)}{\operatorname{p_{\mathbf{Y}}}(\mathbf{y}) \cdot \operatorname{p_{Z}}(z)}} \IEEEnonumber \\
&=2 n + \sum_{\mathbf{y}}  (2^n-1) \cdot \frac{\operatorname{p_{\mathbf{Y}Z}}(\mathbf{y},0)}{2^n-1} \cdot \log{\frac{\operatorname{p_{\mathbf{Y}Z}}(\mathbf{y},0)}{2^n-1}} + \IEEEnonumber \\
&+\operatorname{p_{\mathbf{Y}Z}}(\mathbf{y},1) \cdot \log{\left[\operatorname{p_{\mathbf{Y}Z}}(\mathbf{y},1) \right] } .%
\end{IEEEeqnarray}
From this discussion, we can conclude that the mutual information is identical for all Boolean functions from the class of functions with $N_1=1$ and $N_0=2^n-1$. 

Let $\mathbf{q}=\{q_i\}$, $\mathbf{p}=\{p_i\}$ and $\mathbf{w}=\{w_i\}$, $\forall i \in \{1,2 \ldots, 2^n \}$, such that, $\forall k \in \{0,1,\ldots,n \}$,
\begin{IEEEeqnarray}{Rl}
&q_i=\operatorname{p_{\mathbf{Y}Z}}(\mathbf{y_{i_k}},1)=\frac{(1-p)^{n-k} \cdot p^k}{2^n},  m_{k}=\binom{n}{k} , \IEEEnonumber \\
&\forall i_k \in \{(\sum_{j=0}^{k-1} m_j+1,\sum_{j=0}^{k-1} m_j+2,\ldots,\sum_{j=0}^{k-1} m_j+m_k \},\IEEEnonumber \\
&p_i=\frac{\operatorname{p_{\mathbf{Y}Z}}(\mathbf{y_{i_k}},0)}{2^n-1}=\frac{1-(1-p)^{n-k} \cdot p^k}{(2^n-1) \cdot 2^n}=\frac{w_i}{2^n}. \IEEEnonumber \\
&\Rightarrow \operatorname{MI}(\mathbf{Y},Z) =2 n +  \sum_{i=1}^{2^n}   (2^n-1) \cdot p_i  \cdot \log{p_i} + \sum_{i=1}^{2^n}  q_i \cdot \log{q_i}. \IEEEnonumber
\end{IEEEeqnarray}
\begin{IEEEeqnarray}{Rl}
&\sum_{i=1}^{2^{n}} q_i \cdot \log{q_i} =\sum_{k=0}^{n} \binom{n}{k} \cdot  \frac{(1-p)^{n-k} \cdot p^k }{2^n} \cdot \log{\frac{(1-p)^{n-k} \cdot p^k }{2^n}} \IEEEnonumber \\
&=\frac{-n}{2^n} \cdot \sum_{k=0}^{n} \binom{n}{k} \cdot  (1-p)^{n-k} \cdot p^k + \frac{\log{(1-p)}}{2^n} \cdot \sum_{k=0}^{n} (n-k) \cdot \binom{n}{k} \cdot \IEEEnonumber \\
&\cdot (1-p)^{n-k} \cdot p^k + \frac{\log{p}}{2^n} \cdot \sum_{k=1}^{n} k \cdot \binom{n}{k} \cdot (1-p)^{n-k} \cdot p^{k} \IEEEnonumber \\
&(n-k) \cdot \binom{n}{k}=\frac{n \cdot (n-1)!}{k! \cdot (n-k-1)!}=n \cdot \binom{n-1}{k} \IEEEnonumber \\
&k \cdot \binom{n}{k}=\frac{n \cdot (n-1)!}{(k-1)! \cdot (n-1-k+1)!}=n \cdot \binom{n-1}{k-1}. \IEEEnonumber \\
& \sum_{i=1}^{2^{n}} q_i \cdot \log{q_i} =\frac{-n}{2^n} +\frac{n \cdot (1-p) \cdot \log{(1-p)}}{2^n} \sum_{k=0}^{n-1} \binom{n-1}{k} \cdot  \IEEEnonumber \\
&\cdot (1-p)^{n-1-k} \cdot p^k + \frac{n \cdot p \cdot \log{p}}{2^n} \cdot \sum_{k=1}^{n}  \binom{n-1}{k-1}  \cdot p^{k-1} \cdot \IEEEnonumber \\
&\cdot (1-p)^{n-1-k+1} =-\frac{n}{2^n} -\frac{n}{2^n} \cdot \operatorname{H}(p). \IEEEnonumber \\
&\sum_{i=1}^{2^{n}} \left(2^n -1 \right) \cdot p_i \cdot \log{p_i}=\frac{2^n -1}{2^n} \cdot \left( -n + \sum_{i=1}^{2^n} w_i \cdot \log{w_i}\right).\IEEEnonumber 
\end{IEEEeqnarray}
Let $a=\displaystyle{\frac{1-p}{2^{n-1}}}$ and  $b=\displaystyle{\frac{p}{2^{n-1}}}$. We want to prove that
\begin{IEEEeqnarray}{Rl}
&\operatorname{MI}(\mathbf{Y},Z) \leq 1-\operatorname{H}(p) \Leftrightarrow \sum_{i=1}^{2^n} \left(2^n -1 \right) \cdot w_i \cdot \log{w_i} \leq \IEEEnonumber \\
&\leq \left(-n\right) \cdot \left( n-1\right) + \left(2^n-n \right) \cdot 2^{n-1} \cdot \left( a \cdot  \log{a} + b \cdot \log{b} \right), \label{Toprove} \IEEEnonumber \\
&\text{ where } \operatorname{H}(p)=-p \cdot \log{p} - (1-p) \cdot \log{(1-p)}.%
\end{IEEEeqnarray}
We need to transform the element $(-n) \cdot (n-1)$, from the right side of the inequality, into a sum of the type $x \cdot \log{x}$, such that the number of elements on the right side of the inequality equals that of the left side. That is, we need $2^n \cdot (2^n -1) - 2 \cdot 2^{n-1} \cdot (2^n -n)=(n-1) \cdot 2^n$ elements. That is, we need to find $x$, such that $(n-1) \cdot 2^n \cdot x \cdot \log{x}=(-n) \cdot (n-1) \Leftrightarrow x=\frac{1}{2^n}$.The right hand side sequence has three distinct elements ordered as $a=\frac{1-p}{2^{n-1}} \geq c=\frac{1}{2^{n}} \geq b=\frac{p}{2^{n-1}}$. The left hand side sequence has the elements ordered as $w_{2^n}=\frac{1- p^n}{2^n -1 } \geq  w_{2^n-1}=\frac{1-\left(1-p \right) \cdot p^{n-1}}{2^n -1 }  \geq \ldots \geq w_{i}=\frac{1-\left(1-p \right)^{n-k} \cdot p^k}{2^n -1 }  \geq \ldots \geq w_{1}=\frac{1-\left(1-p \right)^{n}}{2^n -1 }$. Let $\mathbf{X}=[x_1 \enskip x_2 \ldots x_{2^n \cdot \left(2^n-1 \right)}]$ and $\mathbf{Y}=[y_1 \enskip y_2 \ldots y_{2^n \cdot \left(2^n-1 \right)}]$ be equal to
\begin{IEEEeqnarray}{Rl}
&\mathbf{X}=\begin{bmatrix} \underbrace{a \enskip a \ldots a}_{\begin{matrix} \scriptstyle{2^{n-1} \cdot \left(2^n-n \right)} \\ \scriptstyle{\text{ elements}} \end{matrix}} \enskip \underbrace{c \enskip c \ldots c}_{\begin{matrix} \scriptstyle{2^n \cdot \left(n-1\right)} \\ \scriptstyle{\text{ elements}} \end{matrix}} \enskip \underbrace{b \enskip b \ldots b}_{\begin{matrix} \scriptstyle{2^{n-1} \cdot \left(2^n-n \right)} \\ \scriptstyle{\text{ elements}} \end{matrix}} \end{bmatrix}, \IEEEnonumber \\
&\mathbf{Y}=\begin{bmatrix} \underbrace{w_{2^n}}_{\begin{matrix} \scriptstyle{2^n-1} \\ \scriptstyle{\text{ elements}} \end{matrix}} \underbrace{w_{2^n-1}}_{\begin{matrix} \scriptstyle{2^n-1} \\ \scriptstyle{\text{ elements}} \end{matrix}} \ldots \underbrace{w_1}_{\begin{matrix} \scriptstyle{2^n-1} \\ \scriptstyle{\text{ elements}} \end{matrix}} \end{bmatrix}.
\end{IEEEeqnarray}
$\Rightarrow \mathbf{X}$ and $\mathbf{Y}$ are in descending order, which satisfies the first condition of Karamata's theorem \cite{Karamata1932}. Let $g:\mathbb{R_+} \rightarrow \mathbb{R}$, $g(x)=x \cdot \log{x}$. Then, $g$ is a convex function.

\subsubsection{We prove that $w_{2^n} \leq a$} 
$\Leftrightarrow \frac{1-p^n}{2^n-1} \leq \frac{1-p}{2^{n-1}} \Leftrightarrow  (2^n-1) \cdot p -2^{n-1} \cdot p^n \leq 2^{n-1} -1$. Let $f(x):\left[0, \frac{1}{2}\right] \rightarrow \mathbb{R_+}, f(x)= (2^n-1) \cdot x -2^{n-1} \cdot x^n$. $ f^{\prime}(x)=2^n-1 - 2^{n-1} \cdot n \cdot x^{n-1} \geq n-2^{n-1} \cdot n \cdot \frac{1}{2^{n-1}}=0 \Rightarrow f^{\prime}(x) \geq 0, \forall x \in \left[0 \enskip \frac{1}{2}\right], \Rightarrow$ the function $f$ is increasing. Let $x^*$ be the critical point of $f$. $f^{\prime}(x)=0 \Rightarrow (x^*)^{n-1}=\frac{2^n-1}{2^{n-1} \cdot n}\geq \frac{1}{2^{n-1}} \Leftrightarrow x^*\geq \frac{1}{2}$  $\Rightarrow f(x) \leq f\left(\frac{1}{2}\right), \forall x \in \left[0 \enskip \frac{1}{2}\right] \Rightarrow  (2^n-1) \cdot p -2^{n-1} \cdot p^n \leq 2^{n-1} -1\Rightarrow w_{2^n} \leq a$.

Let $\operatorname{SL}_{k}$ and $\operatorname{SR}_{k}$, $\forall k \in \{1,2,\ldots 2^n \cdot(2^n-1) \}$, denote the partial sums computed with the elements of the left-hand sequence of the inequality $($\ref{Toprove}$)$ and with the right-hand one, respectively. Let $K=2^{n-1} \cdot (2^n-n)$. Using the binomial theorem \cite{AS1972}, $2^{n-1} \geq 1+n-1 \Rightarrow  2^{n-1} \cdot \frac{(n-1)}{2^{n-1}-1} \leq 2^{n-1} \leq 2^{n}-1 \Rightarrow $ $ 2^n-1 \leq 2^{n-1} \cdot (2^n-n) \label{Ineqaibi}$. $w_k \leq w_{2^n} \leq a, \forall k \in \{ 2^n, 2^n-1, \ldots, 1 \} \Rightarrow \operatorname{SL}_{k}=\sum_{j=1}^{k} y_j \leq$$ \operatorname{SR}_{k}=\sum_{j=1}^{k} x_j=k \cdot x_1=k \cdot a, \forall k \in \{ 1,2, \ldots, K \} \label{Set1PS}.$

\subsubsection{We prove that $2 \cdot w_{2^n} \leq a +c$} $\Leftrightarrow 2 \cdot \frac{1-p^n}{2^n-1}+  \frac{p}{2^{n-1}} \leq \frac{3}{2^{n}}
\label{IneqUi}$. Let $f(x):\left[0, \frac{1}{2}\right] \rightarrow \mathbb{R_+}, f(x)= 2 \cdot \frac{1-x^n}{2^n-1}+  \frac{x}{2^{n-1}}$. Using the binomial theorem \cite{AS1972}, $\frac{2\cdot n}{2^n-1} \leq 1$ $\Rightarrow \frac{2 \cdot n \cdot x^{n-1}}{2^n-1} \leq \frac{1}{2^{n-1}}, \forall 0 \leq x \leq \frac{1}{2} $. $f^{\prime}(x)=\frac{-2 \cdot n \cdot x^{n-1}}{2^n-1} + \frac{1}{2^{n-1}}$ $\Rightarrow f^{\prime}(x) \geq 0, \forall 0 \leq x \leq \frac{1}{2}$ $\Rightarrow f$ is increasing $\Rightarrow f(x) \leq f\left( \frac{1}{2} \right), \forall 0 \leq x \leq  \frac{1}{2} \Rightarrow 2 \cdot \frac{1-p^n}{2^n-1}+  \frac{p}{2^{n-1}} \leq \frac{3}{2^{n}} \Rightarrow 2 \cdot w_{2^n} \leq a +c$.

\subsubsection{We prove that the inequalities involving the partial sums from Karamata's theorem hold}. If $n=2$, it can be easily verified that $\operatorname{SL}_{K+i} \leq 4 \cdot a + i \cdot c=\operatorname{SR}_{K+i}, \forall i \in \{1,2,3,4\}$. If $n \geq 3$, using the binomial theorem \cite{AS1972}, we have that $K -2^n \cdot (n-1) \geq 1, \forall n\geq 3$; $2 \cdot w_{2^n} \leq a +c \Rightarrow w_j + w_k \leq a +c, \forall j,k \in \{ 2^n,2^n-1,\ldots,1\}$ $\Leftrightarrow y_j + y_k \leq x_{1} +x_{K+1}, \forall j,k \in \{ 2^n \cdot (2^n-1),2^n\cdot (2^n-1)-1,\ldots,1\}$; $ K-i \geq 1, \forall  i \in \{1,2, \ldots, 2^n \cdot (n-1) \}$ $\Rightarrow \operatorname{SL}_{K+i}=\operatorname{SL}_{K-i}+ y_{K-i+1} + \ldots + y_{K} + y_{K+1} + \ldots + y_{K+i} =\operatorname{SL}_{K-i} + (y_{K-i+1} + y_{K+1}) + \ldots + (y_{K}+y_{K+i})$ $\Rightarrow \operatorname{SL}_{K+i} \leq  \operatorname{SR}_{K-i} +i \cdot (x_1+x_{K+1})=\operatorname{SR}_{K+i}, \forall i \in \{1,2, \ldots, 2^n \cdot (n-1) \}$ $\Rightarrow \operatorname{SL}_{K+i} \leq \operatorname{SR}_{K+i}, \forall i \in \{1,2, \ldots, 2^n \cdot (n-1) \} \label{IneqKi}$.

\subsubsection{We prove that $w_1\geq b$}$\Leftrightarrow 2^{n-1}\geq 2^{n-1} \cdot (1-p)^n + (2^n-1) \cdot p$. Let $f(x):\left[0, \frac{1}{2}\right] \rightarrow \mathbb{R_+}, f(x)= 2^{n-1} \cdot (1-x)^n + (2^n-1) \cdot x$. Let $x^*$ be the critical point of $f$. $f^{\prime}(x)=  2^{n-1} \cdot n \cdot (1-x)^{n-1} \cdot (-1) + (2^n-1)$, $f^{\prime \prime}(x)=n \cdot (n-1) \cdot 2^{n-1} \cdot (1-x)^{n-2} \geq 0, \forall x \in \left[ 0 \enskip \frac{1}{2} \right] \Rightarrow f$ is a convex function and $x^*$ is a minimum point $\Rightarrow f(x) \leq f(0)=f(\frac{1}{2})=2^{n-1},\forall x \in \left[ 0 \enskip \frac{1}{2} \right]$ $\Rightarrow  2^{n-1}\geq 2^{n-1} \cdot (1-p)^n + (2^n-1) \cdot p \Rightarrow w_1\geq b$.

\subsubsection{We verify that the final inequalities involving the partial sums from Karamata's theorem hold} $\operatorname{SL}_{2^n \cdot (2^n-1)}=\sum_{i=1}^{2^n} \left(2^n-1 \right) \cdot w_i =2^n-1$. $\operatorname{SR}_{2^n \cdot (2^n-1)}= (n-1) \cdot 2^n \cdot \frac{1}{2^n} + (2^n -n) \cdot 2^{n-1} \cdot \frac{1-p}{2^{n-1}} + (2^n -n) \cdot 2^{n-1} \cdot \frac{p}{2^{n-1}}=2^n-1 \Rightarrow \operatorname{SL}_{2^n \cdot (2^n-1)}=\operatorname{SR}_{2^n \cdot (2^n-1)}$ $\Leftrightarrow \operatorname{SL}_{2^{n}  \cdot (2^n-1) -k} + k \cdot w_1=\operatorname{SR}_{2^{n}  \cdot (2^n-1) -k} + k \cdot b, \forall k \in \{1,2,\ldots, 2^n-1 \} $ $\Leftrightarrow  \operatorname{SL}_{2^{n}  \cdot (2^n-1) -k} =\operatorname{SR}_{2^{n}  \cdot (2^n-1) -k} + k \cdot( b-w_1), \forall k \in \{1,2,\ldots, 2^n-1 \}$ $\Rightarrow \operatorname{SL}_{2^{n}  \cdot (2^n-1) -k}\leq \operatorname{SR}_{2^{n}  \cdot (2^n-1) -k}, \forall k \in \{1,2,\ldots, 2^n-1 \}$.

In  $($\ref{Ineqaibi}$)$, we proved that $2^n-1 \leq$$ 2^{n-1} \cdot (2^n-n) $. $K=2^{n-1} \cdot (2^n-n) $ represents the total number of elements equal to $b$. The partial sum inequalities hold only for $2^n-1$ elements equal to $b$. We need to determine that the remaining number of elements equal to $b$, satisfy the partial sum inequalities. We denote them as $\{\operatorname{SL}_{2^{n}  \cdot (2^n-1) - 2^{n}}, \ldots,\operatorname{SL}_{2^{n}  \cdot (2^n-1) - 2^{n-1} \cdot (2^n-n)+1} \}$ and $\{\operatorname{SR}_{2^{n}  \cdot (2^n-1) - 2^{n}}, \ldots, \operatorname{SR}_{2^{n}  \cdot (2^n-1) - 2^{n-1} \cdot (2^n-n)+1} \}$. 

Let $M=2^{n}  \cdot (2^n-1) -(2^n-1)$. $\operatorname{SL}_{M}=\sum_{j=1}^{M-i} y_j + y_{M-i+1} + \ldots +y_{M} \leq$$ \operatorname{SR}_{M}=\sum_{j=1}^{M-i} x_j + x_{M-i+1}+ \ldots x_{M}, \forall i \in \{1,2, \ldots 2^{n-1} \cdot (2^n-n)-(2^n-1) \}$$\Rightarrow \operatorname{SL}_{M-i} \leq \operatorname{SR}_{M-i} + (b-y_{M-i+1}) + \ldots + (b-y_M) \leq \operatorname{SR}_{M-i},  \forall i \in \{1,2, \ldots 2^{n-1} \cdot (2^n-n)-(2^n-1) \}$$\Rightarrow \operatorname{SL}_{M-i} \leq \operatorname{SR}_{M-i}, \forall i \in \{ 1,2, \ldots, 2^{n-1} \cdot (2^n-n)-(2^n-1) \}$. These sums are well defined, because $M-i \geq 1$, $\forall  i \in \{ 1,2, \ldots, 2^{n-1} \cdot (2^n-n)-(2^n-1) \}$. $\forall i \in \{ 1,2, \ldots, 2^{n-1} \cdot (2^n-n)-(2^n-1) \} \Rightarrow M-i\geq \left[ 2^{n}  \cdot (2^n-1) -(2^n-1) \right]-\left[2^{n-1} \cdot (2^n-n)-(2^n-1) \right]$ $\Leftrightarrow M-i\geq 2^{n}  \cdot (2^n-1) -2^{n-1} \cdot (2^n-n)$.

The first partial sum that does not contain an element equal to $b$ is given by $i=2^{n-1} \cdot (2^n-n)-(2^n-1)$ $\Rightarrow M-i=2^{n}  \cdot (2^n-1) -2^{n-1} \cdot (2^n-n)=K+2^n \cdot (n-1)$. As a result, $\operatorname{SL}_{K+2^n \cdot (n-1)} \leq \operatorname{SR}_{K+2^n \cdot (n-1)}$,  which we also proved in $($\ref{IneqKi}$)$. In conclusion, all the conditions in Karamata's theorem are satisfied. This yields $\sum_{i=1}^{2^n \cdot (2^n-1)} g(y_i) \leq \sum_{i=1}^{2^n \cdot (2^n-1)} g(x_i) \Leftrightarrow \operatorname{MI}(\mathbf{Y},Z) \leq 1-\operatorname{H}(p)$.

%
%

Following the above reasoning, the same result holds, for Boolean functions that have one element equal to $0$ in their output table and the rest are equal to $1$, that is $N_1=2^n-1$ and $N_0=1$.

\subsection{Boolean functions from the classes $3$ and $4$ of Theorem \ref{ThCK}} 
\label{Case2}
For any $r \in \{ 1,2,\ldots, n-1\}$, let $N_1=2^{n-r}$ and $\mathbf{Y^{(k)}}=[\mathbf{Y_r^{(k)}} \enskip \mathbf{Y_{n-r}^{(k)}}]$, $\forall k \in \{1,2, \ldots, 2^n \}$, and $\mathbf{X^{(i)}}=[\mathbf{X_r} \enskip \mathbf{X_{n-r}^{(i)}}]$, $\forall i \in \{1,2, \ldots, 2^{n-r} \}$, such that $\mathbf{X^{(i)}}$ $\in \{ [\mathbf{X_r} \enskip 0  \enskip  0 \ldots 0  \enskip  0], [\mathbf{X_r}  \enskip 0  \enskip 0 \ldots 0 \enskip  1], \ldots, [\mathbf{X_r}  \enskip 1  \enskip 1 \ldots 1 \enskip  1] \}$. The output table of the Boolean function has $N_1=2^{n-r}$ number of ones, such that these values correspond to the vector of inputs $\mathbf{X^{(i)}}$ $\in \{ [\mathbf{X_r} \enskip 0  \enskip  0 \ldots 0  \enskip  0], [\mathbf{X_r}  \enskip 0  \enskip 0 \ldots 0 \enskip  1], \ldots, [\mathbf{X_r}  \enskip 1  \enskip 1 \ldots 1 \enskip  1] \}$, where $\mathbf{X_r}$ is fixed. The rest of the output values are zeros.

From the properties of the binary symmetric channel, we have that $ \operatorname{p}(\mathbf{Y^{(k)}},\mathbf{X^{(i)}})= \operatorname{p}(\mathbf{Y^{(k)}_{r}},\mathbf{X_{r}}) \cdot  \operatorname{p}(\mathbf{Y^{(k)}_{n-r}},\mathbf{X_{n-r}^{(i)}}), \forall k \in \{1,2, \ldots, 2^n \}$. According to Lemma \ref{Lem0}, $\sum_{i=1}^{2^{n-r}} \operatorname{p}(\mathbf{Y^{(k)}_{n-r}},\mathbf{X_{n-r}^{(i)}})=\frac{1}{2^{n-r}}$. Let $q_k=\operatorname{p_{\mathbf{Y}Z}}(\mathbf{Y^{(k)}},1)=\sum_{i=1}^{2^{n-r}}  \operatorname{p}(\mathbf{Y^{(k)}},\mathbf{X^{(i)}})=\sum_{i=1}^{2^{n-r}} \operatorname{p}(\mathbf{Y^{(k)}_{r}},\mathbf{X_{r}}) \cdot  \operatorname{p}(\mathbf{Y^{(k)}_{n-r}},\mathbf{X_{n-r}^{(i)}})= \frac{\operatorname{p}(\mathbf{Y^{(k)}_{r}},\mathbf{X_{r}}) }{2^{n-r}}$. Let $p_k=\operatorname{p_{\mathbf{Y}Z}}(\mathbf{Y^{(k)}},0)=\operatorname{p_{\mathbf{Y}}}(\mathbf{Y^{(k)}}) -\operatorname{p_{\mathbf{Y}Z}}(\mathbf{Y^{(k)}},1)=\frac{1}{2^n} - \operatorname{p_{\mathbf{Y}Z}}(\mathbf{Y^{(k)}},1), \forall k \in \{1,2, \ldots, 2^n \}$. For any $k \in \{1,2, \ldots, 2^n \}$, the total number of $\mathbf{Y^{(k)}}=[\mathbf{Y_r^{(k)}} \enskip \mathbf{Y_{n-r}^{(k)}}]$ that have the same $\mathbf{Y_r^{(k)}}$ is equal to $N_1=2^{n-r}$. This produces a number of $N_1=2^{n-r}$ identical probability mass values, $ q_k=\displaystyle{\frac{\operatorname{p}(\mathbf{Y^{(k)}_{r}},\mathbf{X_{r}}) }{2^{n-r}}}$ and $N_1=2^{n-r}$ identical probability mass values, $p_k=\displaystyle{\frac{1}{2^n}}-q_k$. Let the vectors $\mathbf{v}=[v_1 \enskip v_2 \ldots v_{2^r}]$ and $\mathbf{t}=[t_1 \enskip t_2 \ldots t_{2^r}]$ denote the distinct values of the vectors $\mathbf{q}=[q_1 \enskip q_2 \ldots q_{2^n}]$ and $\mathbf{p}=[p_1 \enskip p_2 \ldots p_{2^n}]$, respectively.\begin{IEEEeqnarray}{Rl}
\operatorname{MI}(\mathbf{Y},Z)=2n+2^{n-r} \sum_{i=1}^{2^r} t_i \cdot \log{\frac{t_i}{2^n-2^{n-r}}} + v_i \cdot \log{\frac{v_i}{2^{n-r}}} \IEEEnonumber
\end{IEEEeqnarray}

For any $\mathbf{X_r} \in \{0,1 \}^r$ fixed, there exists: one vector, that is $m_0=1$, $\mathbf{Y_r^{(i_0)}} \in \{0,1 \}^r$, such that $\mathbf{Y_r^{(i_0)}}=\mathbf{X_{r}}$, a number $m_1=\binom{r}{1}$ of the vectors $(\mathbf{Y_r^{(i_1)}})$, $\forall i_1 \in \{m_0+1,m_0+2,\ldots,m_0+m_1 \}$, such that $(\mathbf{Y_r^{(i_1)}})$ differ from $\mathbf{X_r}$ in one position and a number $m_j=\binom{r}{j}$ of the vectors $(\mathbf{Y_r^{(i_j)}})$, $\forall i_j \in \{(m_0+\ldots+m_{j-1})+1,(m_0+\ldots+m_{j-1})+2,\ldots,(m_0+\ldots+m_{j-1})+m_j \}$, such that $(\mathbf{Y_r^{(i_j)}})$ differ from $\mathbf{X_r}$ in $j$ positions, $\forall j \in \{0,1,2, \ldots r\}$. As a result, we obtain
\begin{IEEEeqnarray}{Rl}
&\operatorname{p}(\mathbf{Y^{(i_j)}_{r}},\mathbf{X_{r}})=\frac{(1-p)^{r-j} \cdot p^j}{2^r}, m_j=\binom{r}{j} , \forall j \in \{0,1, \ldots,r\} \IEEEnonumber \\
&\forall i_j \in \{(m_0+\ldots+m_{j-1})+1,\ldots,(m_0+\ldots+m_{j-1})+m_j \IEEEnonumber \\
&v_i=\frac{(1-p)^{r-j} \cdot p^{j}}{2^r \cdot 2^{n-r}},  t_i=\frac{1-(1-p)^{r-j} \cdot p^{j}}{2^n} \IEEEnonumber \\
&\Rightarrow \operatorname{MI}(\mathbf{Y},Z)=2r+ \sum_{i=1}^{2^r} (2^{n-r} \cdot t_i) \cdot \log{\frac{(2^{n-r} \cdot t_i)}{2^r-1}} + \IEEEnonumber \\
&+ (2^{n-r} \cdot v_i) \cdot \log{(2^{n-r} \cdot v_i)} \leq 1 - \operatorname{H}(p).%
\end{IEEEeqnarray}
The last inequality represents the result proved for Boolean functions from the classes $1$ and $2$, with $n=r$. Equality is obtained for $r=1$, that is for the dictatorship function. If $r=1 \Rightarrow N_1=2^{n-1}, N_0=2^{n-1}, \mathbf{v}=\left[\frac{1-p}{2^n} \enskip \frac{p}{2^n}\right]$ and $\mathbf{t}=\left[\frac{p}{2^n} \enskip \frac{1-p}{2^n} \right]$ $\Rightarrow \operatorname{MI}(\mathbf{Y},Z)=1 - \operatorname{H}(p)$.

Following the above reasoning, the same result holds, for Boolean functions that have $N_0=2^{n-r}$ elements equal to $0$ in their output table and the rest are equal to $1$, that is $N_1=2^n-2^{n-r}=2^{n-r} \cdot (2^r-1)$, $\forall r \in \{1,2, \ldots, n-1\}$. These Boolean functions satisfy an additional condition: the $0$ values from the output table correspond to the input vectors $\mathbf{X^{(i)}}=[\mathbf{X_r} \enskip \mathbf{X_{n-r}^{(i)}}]$ $\in \{ [\mathbf{X_r} \enskip 0  \enskip  0 \ldots 0  \enskip  0]$,$ [\mathbf{X_r}  \enskip 0  \enskip 0 \ldots 0 \enskip  1]$, $\ldots, [\mathbf{X_r}  \enskip 1  \enskip 1 \ldots 1 \enskip  1] \}$, where $\mathbf{X_r}$ is fixed, $\forall i \in \{1,2,\ldots,2^{n-r} \}$.

\section{Conclusions}
\label{Conc}
In this study, we proved the Courtade-Kumar conjecture, for certain subclasses of Boolean lex functions, for all dimensions, $\forall n \geq 2$, and for all values of the error probability, $\forall 0 \leq p \leq \frac{1}{2}$. We provided an algebraic proof using Karamata's theorem as our main tool. We brought further improvement in the effort to establish this conjecture in its most general form. Our novelty lied in showing that, for several subclasses of Boolean lex functions, the conjecture holds for all dimensions, $\forall n \geq 2$, and for all values of the error probability, $\forall 0 \leq p \leq \frac{1}{2}$. We have tried to apply Karamata's theorem to other types of Boolean functions, in order to solve the conjecture in its most general form. However, we have been unsuccesful in both applying the theorem directly to the mutual information inequality and in finding a suitable algebraic transformation of the original inequality into an expression that can be proved with Karamata's theorem. The majorazation condition from this theorem cannot be verified. 
\section{Acknowledgments}
We would like to thank Thomas Courtade for helpful discussions on lex functions and for indicating two articles which employ Karamata's theorem and its extension, Schur convexity, namely an earlier version of the preprint \cite{Kindler2016} and \cite{Kumar2013}, respectively.  
\bibliographystyle{IEEEtran}
\bibliography{Conjecture1}

%








\end{document}